# High-Frequency and Microwave Magnetic Properties of $Ni_{0.5}Zn_{0.5}Fe_2O_4$ Spinel Ferrite Ceramics

M. Kempa,[1a] V. Bovtun,[1] V. Kukhar,[2] S. Solopan,[2] A. Belous,[2] O. V'yunov,[2] Y. Yakymenko,[3] S. Kamba[1]

[1]Institute of Physics of the Czech Academy of Sciences, Na Slovance 2, 182 00 Prague, Czech Republic

[2]Department of Solid State Chemistry, V.I. Vernadsky, Institute of General and Inorganic Chemistry, NAS of Ukraine, 32/34 Palladina Ave, Kyiv, Ukraine

[3]Department of Microelectronics, NTUU "Igor Sikorsky Kyiv Polytechnic Institute", Prospect Beresteiskyi 37, 03056 Kyiv, Ukraine

[a]Corresponding author: kempa@fzu.cz

## Abstract

Magnetic properties of the $Ni_{0.5}Zn_{0.5}Fe_2O_4$ (NZF) spinel ferrite ceramics were studied over a broad frequency range (1 MHz – 50 GHz). Between 10 MHz and 2 GHz, strong temperature-dependent resonance-like magnetic permeability dispersion was observed and attributed to the magnetic domain-wall dynamics. It is responsible for the high magnetic losses, absorption and shielding ability of NZF, and provides high nonlinearity and tunability of the permeability under a weak magnetic field. The attenuation constant of NZF is comparable to those of dielectric-conductor composites and giant permittivity materials. In the microwave range (2–50 GHz), three magnetic excitations dependent on a weak magnetic field were revealed and related to magnons. The lowest-frequency magnon (<10 GHz) is attributed to the natural ferromagnetic resonance, two others are excited between 28 and 44 GHz. Interaction of the magnons and magneto-dielectric resonance modes with electromagnetic waves provides high absorption and shielding efficiency in the GHz range, including 5G and 6G communication frequencies.

## 1. **Introduction**

Development of materials providing effective absorption and shielding of microwave (MW) electromagnetic (EM) radiation becomes a more and more actual problem [1–3]. It is related to the increase of EM pollution due to extensive development of the wireless communications and MW electronic systems [4,5]. Also, MW absorbing and shielding materials are widely used in the defence technologies [6]. To absorb EM radiation, materials should be characterized by high magnetic and dielectric losses in the high-frequency (HF) and/or MW ranges [7,8], i.e. from 1 MHz to 50 GHz, and even to 100 GHz, depending on the application. During the last twenty years, ferrite-based materials (ceramics, nanocrystals, composites, films) were studied and developed as effective absorbing materials [9,10]. Study of material properties (magnetic and dielectric, dynamic parameters) in relation to the technology, structure and size (thickness) of the absorbing layer enables the optimization of the chemical composition for the required frequency range [11–13].

Spinel ferrites (see the comprehensive review [14]) have been used as absorbing materials for a long time, due to their high MW magnetic losses near the natural ferromagnetic resonance (FMR) frequency. The FMR frequency depends on the magnetization value (or magnetic anisotropy). The frequency of the natural FMR (without any bias field) in the previously not magnetized ceramics is also related to the exchange interaction between neighbouring grains [15] and therefore it depends on the grain structure. Nickel-zinc spinel ferrites $Ni_{1-x}Zn_xFe_2O_4$ are characterized by high magnetization and low coercive field [16,17]. Additionally, an increase of the Zn concentration results in an essential decrease of the paramagnetic-ferrimagnetic transition temperature (which in pure $NiFe_2O_4$ is 850 K [18,19]) and, consequently, allows to control the natural FMR frequency.

Absorbing materials of various kind and morphology have been prepared (by various technologies) on the base of nickel-zinc spinel ferrites and studied: ceramics, nanoparticles, composites, etc. [20–26]. Ferrite nanofibers prepared by the electrospinning method [20], composites of the ferrite nanoparticles and carbon nanotubes [21], composites of the ferrite nanoparticles and conductive polymers [22] were found to exhibit increased absorption. Influence of the cation substitution in sublattices on the absorbing properties was studied [23,26].

Development of the 5G and future 6G communication systems working at ~30 GHz and ~60 GHz, respectively, requires absorbing and shielding materials for the corresponding frequency range. Contemporary absorbing materials were studied mostly below 20 GHz, and their absorbing properties are mainly based on the mechanisms of dielectric and magnetic losses in this frequency range which are well summarized in [23]: conductivity, dielectric relaxations of different kinds, dynamics of the magnetic domain walls, magnetic hysteresis, eddy current effects, spin rotation, FMR. Above 20 GHz, the contributions of all these mechanisms are essentially reduced. Microwave absorption mechanisms were also recently analysed for the U-type hexaferrite $Sr_4CoZnFe_{36}O_{60}$ ceramics above 20 GHz [27,28]. This ceramics crystallizes in the $R\bar{3}m$ space group and its structure can be described as the stacking sequence of three main structural blocks $(RSR^*S^*TS^*)_3$. Detailed crystal structure of $Sr_4Co_2Fe_{36}O_{60}$ was published by Honda et al. [29], details of the preparation and results of the characterization are described in [27]. Most probably, interaction of the microwaves with the higher-frequency magnons (above that of the natural FMR) revealed in the U-type hexaferrite between 20 and 50 GHz both in the conical and collinear magnetic phases [28] will contribute to the absorption and shielding also in other similar materials, including spinel ferrites with the collinear magnetic phase.

For a better understanding and successful utilization of all potential absorption mechanisms, a broadband study of the material's magnetic properties is necessary. To our knowledge, such studies (up to 50 GHz) are scarce and not systematic. This motivates us to study and analyse the magnetic properties of the well-known absorbing material, $Ni_{0.5}Zn_{0.5}Fe_2O_4$ (NZF) spinel ferrite ceramics, in the broad frequency range from 1 MHz to 50 GHz. According to the literature, NZF is ferrimagnetic below ~550 K [30].

## 2. Material processing and experimental methods.

The powder precursors of NZF with a spinel structure were obtained by the method of the sequential precipitation of hydroxides [31]. For synthesis, aqueous solutions of iron $Fe(NO_3)_3$, nickel $Ni(NO_3)_2$ and zinc $Zn(NO_3)_3$ nitrates, and a 1 molar aqueous solution of sodium hydroxide as the precipitator were used. The precipitation of salts was carried out sequentially at *pH* values selected individually for each metal cation. At the first stage, the bi-distilled water was poured in a reactor, the *pH* value was brought to 4-4.5 using nitric acid ($HNO_3$) or sodium hydroxide (NaOH), then the aqueous solution of the $Fe(NO_3)_3$ salt and the precipitator (solution of NaOH) were added as droplets with a continuous stirring. After the precipitation, the *pH* value of the mother solution was brought to 7.0-7.2, solutions of the $Zn(NO_3)_3$ salt and precipitator were being added sequentially as droplets with continuous stirring. After the previous precipitation, the *pH* value of the mother solution was raised to 8.5-8.7 and sequentially the droplets of the $Ni(NO_3)_2$ salt and precipitator were being added.

The control of precipitator addition was carried out by the automatic titration block with a *pH*-meter. It made it possible to obtain the precipitates of any component under conditions of constant *pH* values. The *pH* value of the mother solution was in the preset range and regulated by salt and alkali aqueous solutions. The speed of the precipitating solution supply was regulated by a peristaltic pump. Precipitation of 100 g of the final product was carried out for 4 hours under intense stirring. After precipitation of all components, the resulting suspension was heated to 80 °C for 1 hour. The precipitate was filtered and washed with a bi-distilled water, until the concentration of sodium ions decreased to 0.1 mg/ml, and then dried at 110-120 °C. The precipitates were calcinated in an air atmosphere at 500-800 °C. Then the NZF ceramics was sintered in an air atmosphere at 1350 °C for two hours.

X-ray diffraction (XRD) studies were performed using the DRON-4 diffractometer (Cu Kα radiation). For the phase characterization, the JCPDS database was used (JCPDS Card Number 52-0279). Crystallographic lattice parameters of the single-phase product were calculated by the Rietveld method using the FullProf software package. The morphology of the NZF ceramics was analysed using the JEM-1230/JEM-1400 transmission electron microscope (TEM) and the Tescan Mira 3 LMU scanning electron microscope (SEM) using a backscattered electron diffraction detector (BSE). The chemical composition was analysed by energy dispersive X-ray (EDX) microanalysis using a conductive gold film coated sample.

The average grain size was determined using both TEM and analysis of diffraction peaks broadening in the XRD patterns. To estimate the coherent scattering region, broadening of the (022) and (044) reflections was analyzed. In order to account for influence of the lattice strain and crystal imperfections on the peaks' broadening, the Williamson–Hall method was employed using the full width at half maximum (FWHM). [32,33].

We used different setups for the high-frequency (HF) and microwave (MW) experiments. The HF complex magnetic permeability $\mu^* = \mu' + i\mu''$ was measured in a reflection coaxial setup from ~1 MHz to ~2 GHz [27] using a 7 mm rigid coaxial line. A ring-shaped sample (toroid) without electrodes with an inner diameter of 3.1 mm, outer diameter of 7.0 mm and a thickness of 1.3 mm was inserted between the inner and outer conductors of an Agilent 16454A magnetic material test fixture (a short-circuited 8-mm coaxial measuring cell, see Fig. 1a) ensuring no electric contact between the sample and both coaxial conductors. The gaps between the sample and the coaxial conductors allow us to neglect the capacitance contribution to the measured impedance and calculate permeability from the inductive contribution, accounting for the sample and coaxial fixture sizes [34]. The 7 mm rigid coaxial line provides precise values of the measured impedance and calculated complex permeability (1-3 % errors). The complex input impedance of the coaxial line with a sample was recorded by the computer-controlled spectrometer consisting of the (a) Agilent 4291B RF impedance analyser and Sigma Systems M18 temperature chamber with liquid nitrogen cooling from ~400 K to ~100 K with a temperature rate of 1 K/min. The temperature-frequency dependence of complex permeability was calculated from the recorded impedance data. Measurements with the applied magnetic bias field were performed at room temperature (RT).

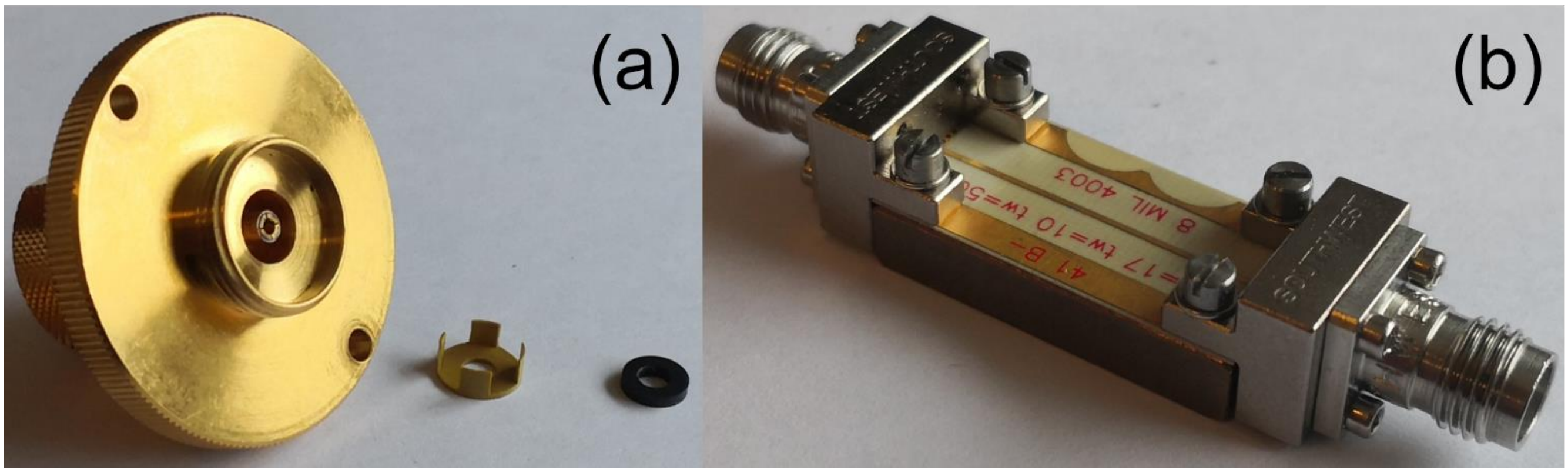

Fig. 1. Photos of (a) the magnetic material test fixture, and (b) the microstrip line, used in our experiments.

MW measurements (100 MHz – 50 GHz) were performed at RT with the Agilent E8364B network analyser using the Southwest microstrip line (MSL) test fixture [35] (Fig. 1b). The MSL dimensions are: length 25.4 mm, width 12.7 mm, thickness 0.20 mm, microstrip line width 0.43 mm.

The same sample (used in the HF experiment) was placed onto the MSL, covering partially the microstrip line, and mechanically pressed down for better contact. A full set of scattering *S*-parameters ($S_{11}$, $S_{22}$, $S_{21}$, $S_{12}$, both magnitude and phase) of the MSL loaded with a sample was recorded in the transmission setup [28,36]. The $S_{21}$ and $S_{11}$ amplitudes are considered as transmission and reflection coefficients *Tr* and *R*, respectively. Reflection loss *RL* is an amplitude of the $S_{11}$ parameter measured in the reflection setup [37,38]) using a "short" calibration standard. *Tr* and *RL* are presented and analysed as most reliable and useful MW parameters [28,36,37,38]. A precise Southwest MSL test fixture with 2.4 mm coaxial connectors [35] assures propagation of the only quasi-TEM mode with transversal electric and magnetic components of the MW electromagnetic field, without any resonance below 50 GHz. After calibration of the test fixture with phase-stable connecting coaxial cables, the high transmission ($Tr = 0 \pm 0.2$ dB), low reflection coefficients ($R \leq -30$ dB) and high reflection loss ($RL = 0 \pm 0.2$ dB) were achieved in the whole frequency range. The influence of the weak bias magnetic field *H* (up to 700 Oe) on the magnetic permeability, transmission and reflection loss spectra was studied at RT in the coaxial and MSL setups [28,36].

## 3. Structure

NZF nanoparticles, synthesized after the heat treatment at 800 °C, are crystalline, single-phase, and have a cubic spinel structure with the $Fd\bar{3}m$ space group (SG). The nanoparticles are nanosized and weakly agglomerated (see the TEM image in Fig. 2a), with an average size, calculated using the mathematical modelling [39], ranging from 30 to 55 nm (Fig. 2b). The lattice parameter a = 8.397(1) **Å** is consistent with previously reported value [40]. According to the Williamson–Hall analysis, the average crystallite size is $D_{XRD}$ = 35 nm, and the microstrain is $\epsilon = 8 \times 10^{-4}$. Full-profile XRD analysis (Rietveld) of the NZF ceramic sintered from nanopowders at 1350 °C revealed diffraction patterns matching the XRD-4 reference card (ICDD PDF-4 No. 96-900-9921), with a refined lattice parameter of a = 8.3953(3) Å. In the spinel structure with the SG $Fd\bar{3}m$, cations occupy the Wyckoff positions 16d (½, ½, ½) and 8a (⅛, ⅛, ⅛). Consequently, the only positional parameter requiring refinement is the oxygen coordinate at the 32e site (x, x, x), which was determined to be x = 0.257(1) for the NZF ceramic. The grain size of the NZF ceramic is about 10 μm (Fig. 2d).

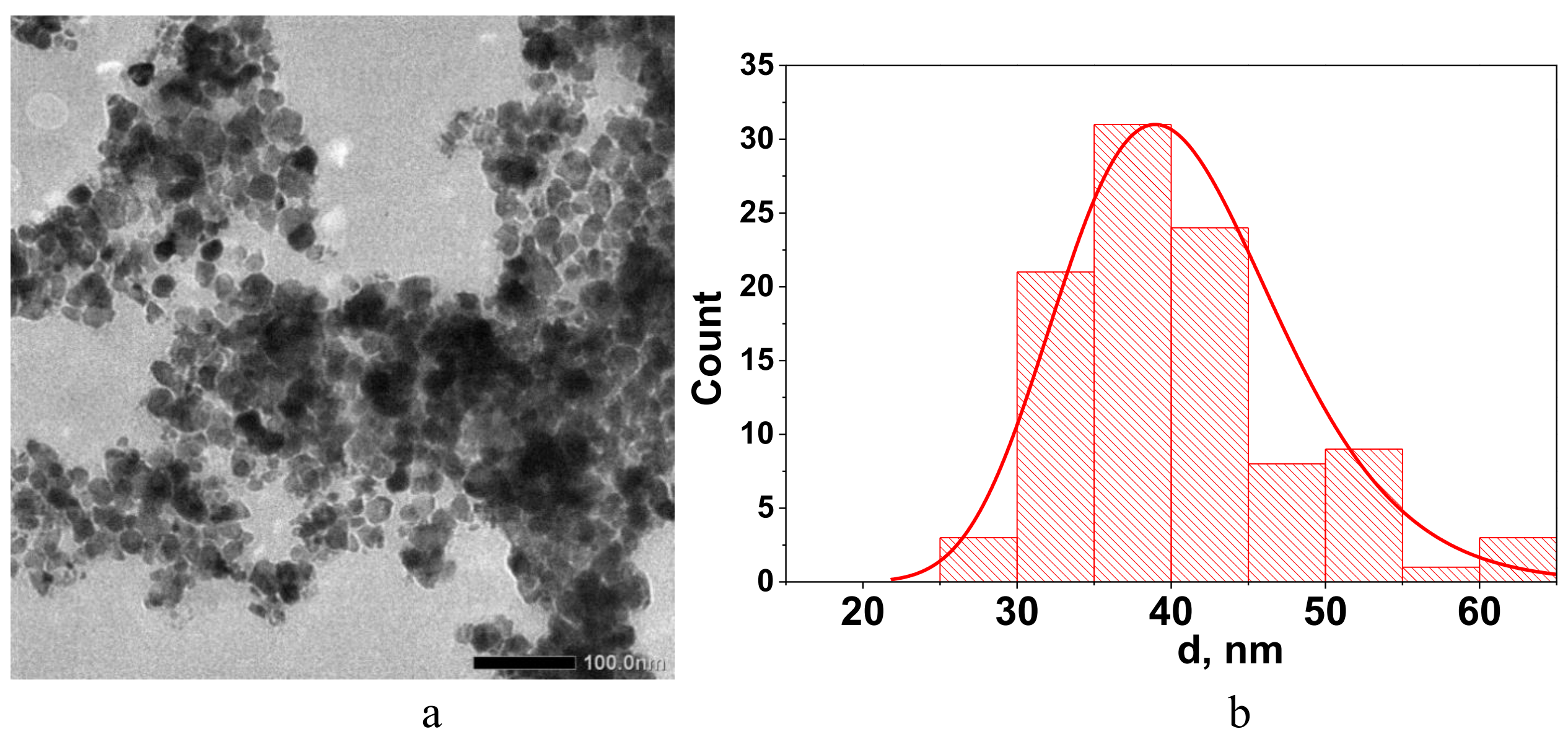

a b

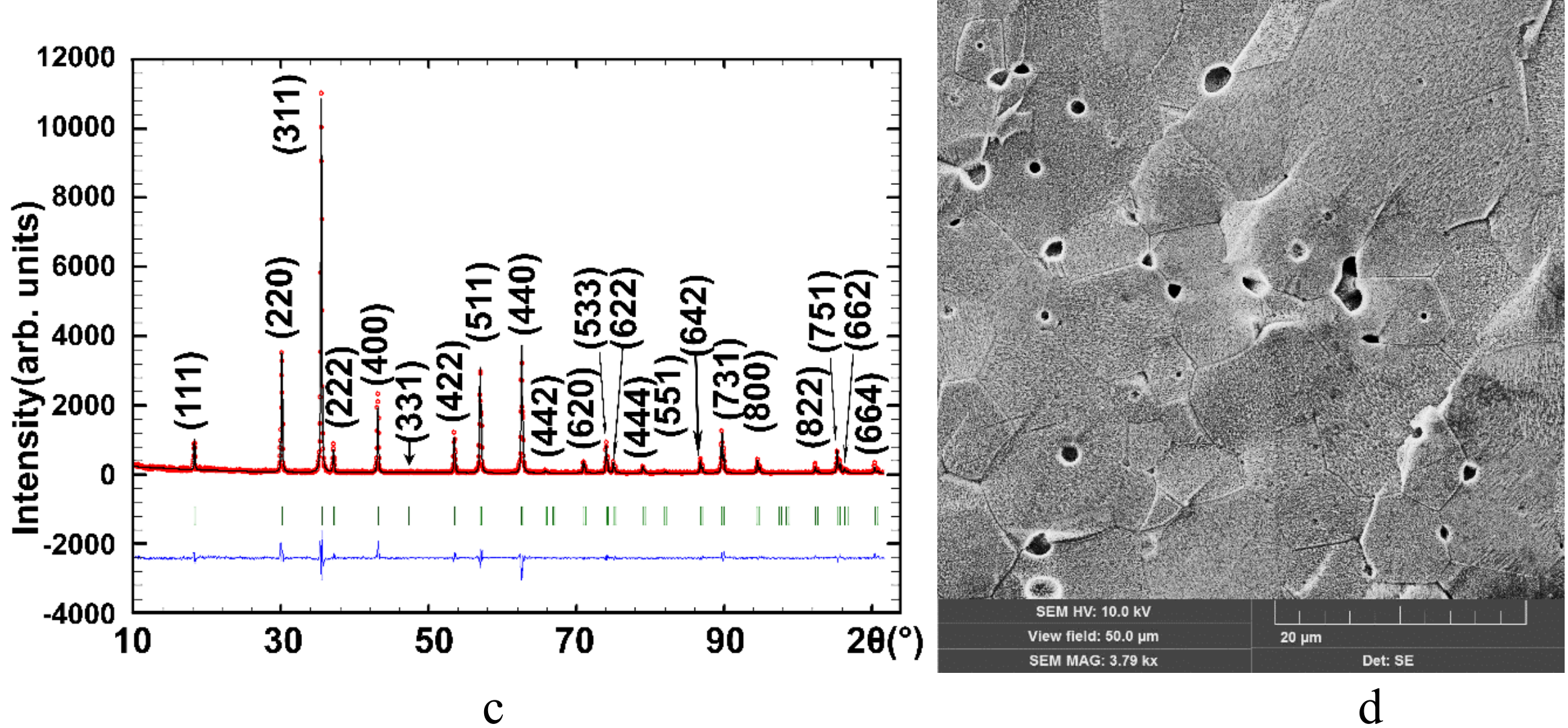

Fig. 2. (a) Morphology and (b) particle size distribution of NZF nanoparticles; (c) X-ray diffraction patterns and (d) microstructure of ceramics sintered from these nanoparticles.

Analysis of the results from Electron Backscattering Diffraction (EBSD) and Energy-Dispersive X-ray Spectroscopy (EDX) of various regions of the sample indicate high uniformity of the ceramic material (Fig. 3). Only the main elements (Ni, Zn, Fe) are observed at different surface points.

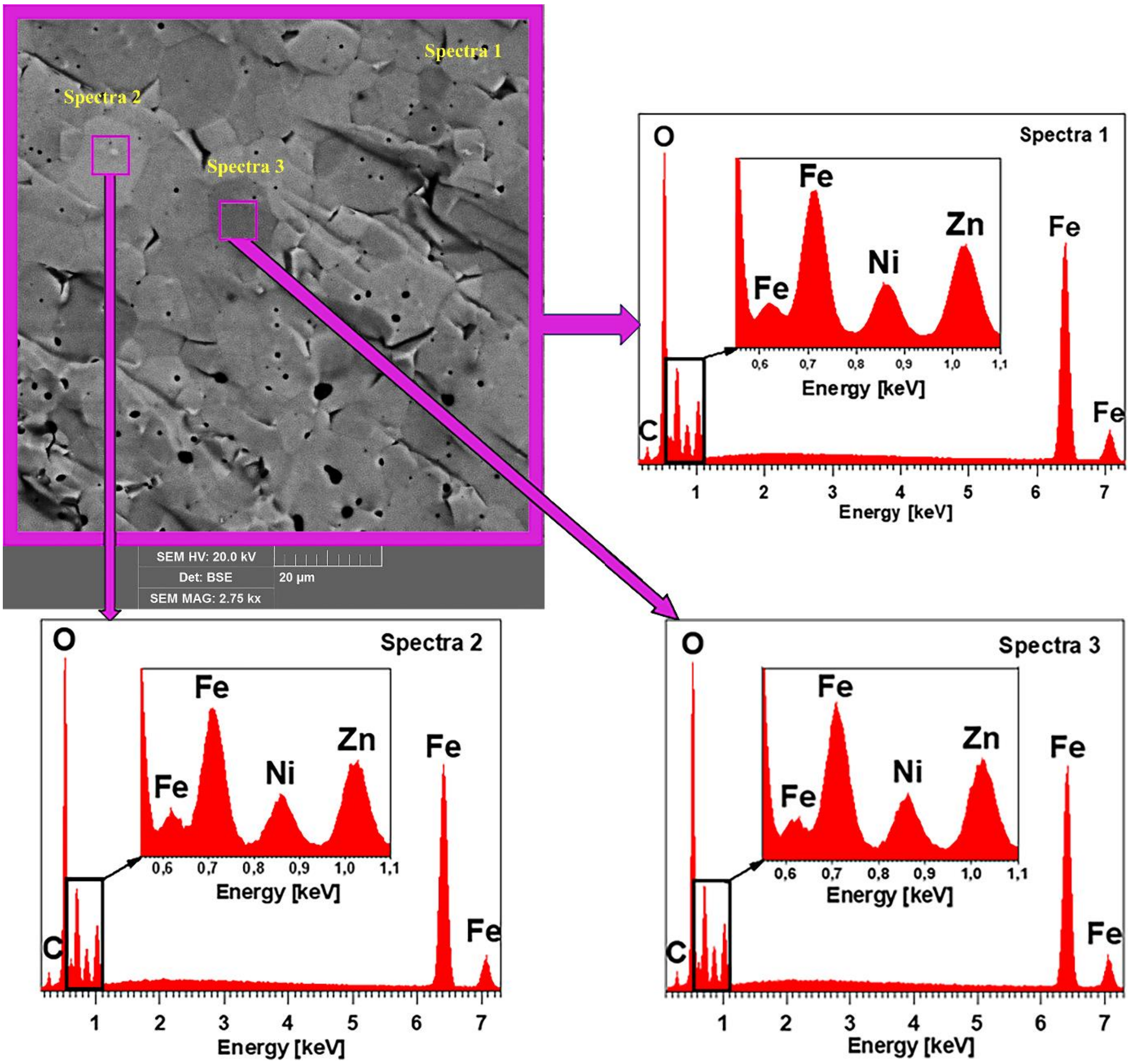

Fig. 3. BSE image of $Ni_{0.5}Zn_{0.5}Fe_2O_4$ ceramic with EDX spectra acquired from different regions (Spectra 1, 2, and 3).

### 4. High-frequency magnetic permeability.

Temperature and frequency dependences of the HF complex magnetic permeability of NZF are shown in Fig. 4. The resonance-like permeability dispersion takes place between 1 MHz and 1.8 GHz at all temperatures (Fig. 4b). A diffuse character of the dispersion could be explained by the presence of grains and multidomain magnetic structure in the previously not magnetized ceramics [14]. As a result of the dispersion, $\mu'$ reduces to the level of ~1 at high temperatures and 2 GHz.

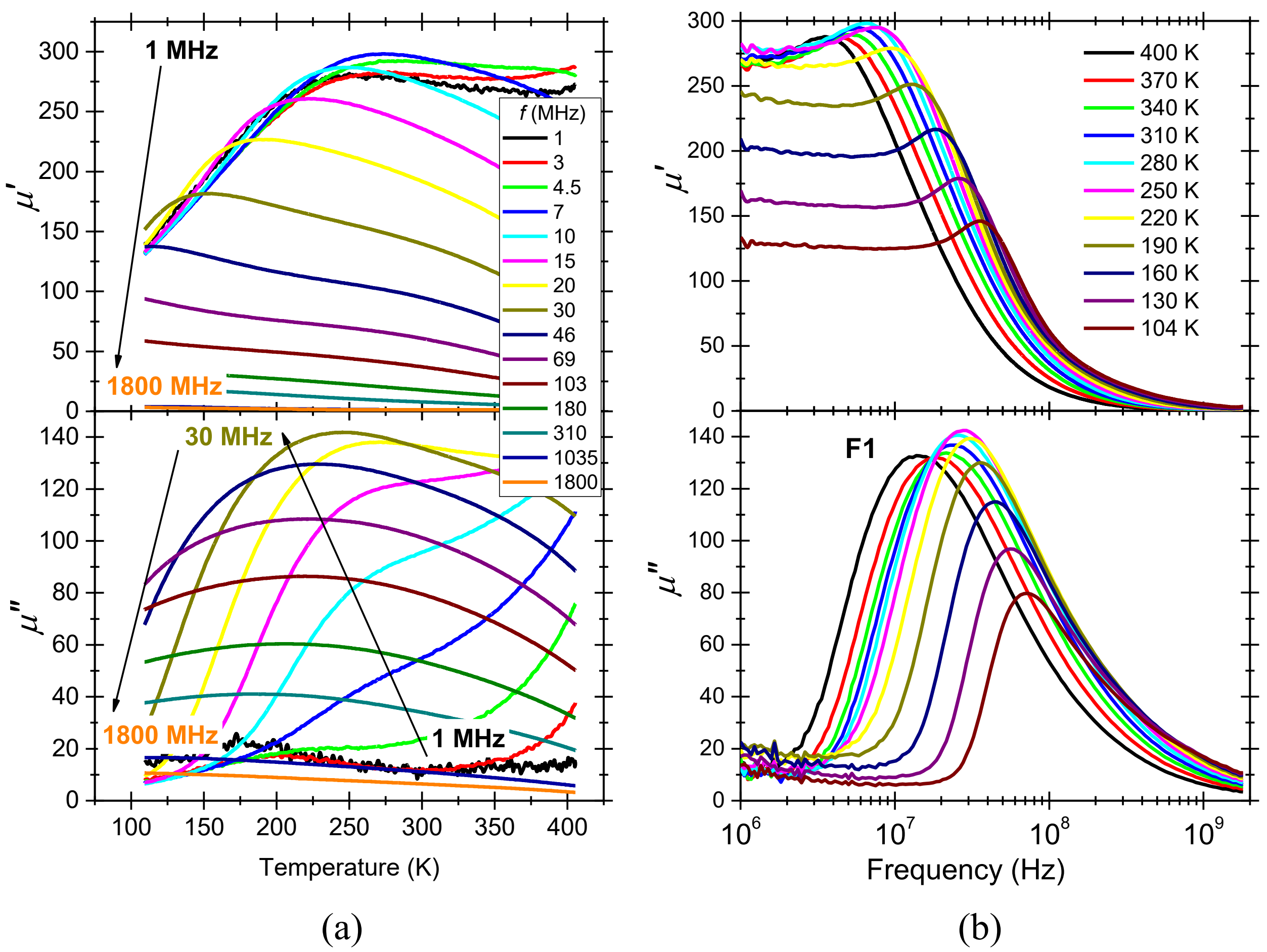

Fig. 4. (a) Temperature and (b) frequency dependences of the HF complex magnetic permeability $\mu^*(T, f)$ of NZF. All results were obtained during cooling in the temperature chamber. The loss maxima $\mu''(f)$ correspond to the F1 magnetic excitation.

The mean resonance frequency $f_1$ corresponding to the loss maxima $\mu''(f)$ is related to the F1 magnetic excitation (marked in Fig. 4b). The $f_1$ decreases with increasing temperature from 70 MHz to 10 MHz. Its dependence is relatively weak and monotonic above 100 K (Fig. 5), with a small anomaly (change of the slope) near 225 K**.** This indicates that the F1 magnetic excitation is not a temperature activated one. Accounting for the $f_1$ frequency range, the F1 excitation can be attributed to dynamics of the domain walls, like in other ferrites [27, 41]. Its temperature dependence reflects the evolution of the domain structure. The low frequencies of the F1 excitation can be caused by the location of domain walls in the NZF ceramics rather inside the grains than on the grain boundaries [42, 43]. Consequently, the domain wall dynamics is rather soft compared to other ferrites, whose domain walls are located on the grain boundaries. The RT loss maximum $\mu''(f)$ is observed at ~20 MHz, similar to that of the NZF nanoparticles [37].

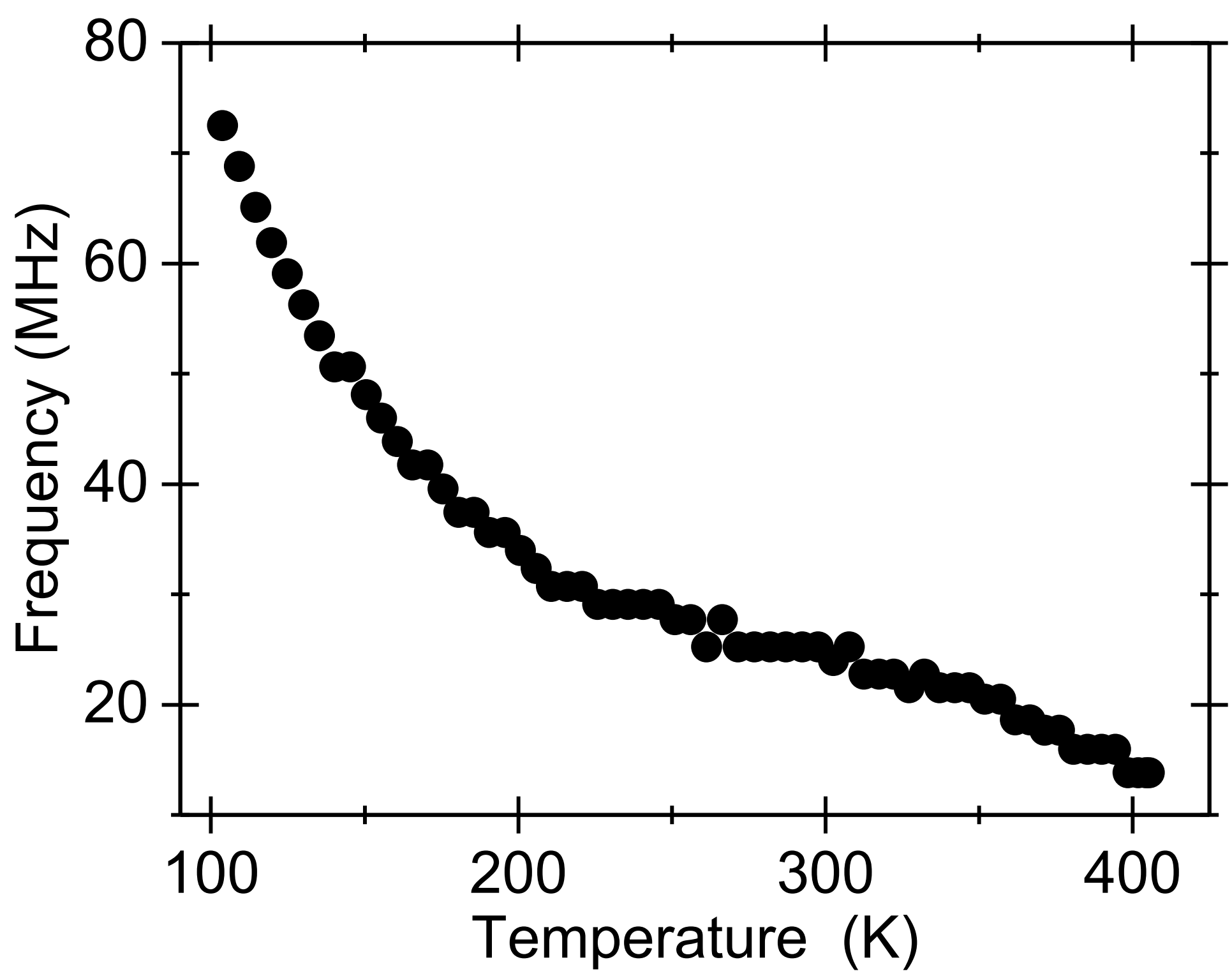

Fig. 5. Temperature dependence of $f_1$ - mean frequency of the F1 excitation.

An influence of the weak bias magnetic field $H$ (up to 700 Oe) on the HF complex magnetic permeability of NZF studied at RT is shown in Fig. 6a. The $\mu'(f)$ maximum decreases from ~280 to ~170, the $\mu''(f)$ maximum decreases from ~130 to ~70. The frequency of the $\mu''(f)$ maximum shifts from ~20 MHz to ~45 MHz. For quantitative analysis of the magnetic nonlinearity, we use a simple method developed for description of the giant magnetoimpedance effect [27,44]: comparison of the permeability $\mu'$ and loss $\mu''$ measured at zero and maximum bias fields ($H_0$ and $H_{max}$) applied in the experiment:

$$n'(\%) = 100\% * \frac{[\mu'(H_0) - \mu'(H_{max})]}{\mu'(H_{max})} \quad (1)$$

$$n''(\%) = 100\% * \frac{[\mu''(H_0) - \mu''(H_{max})]}{\mu''(H_{max})} \quad (2)$$

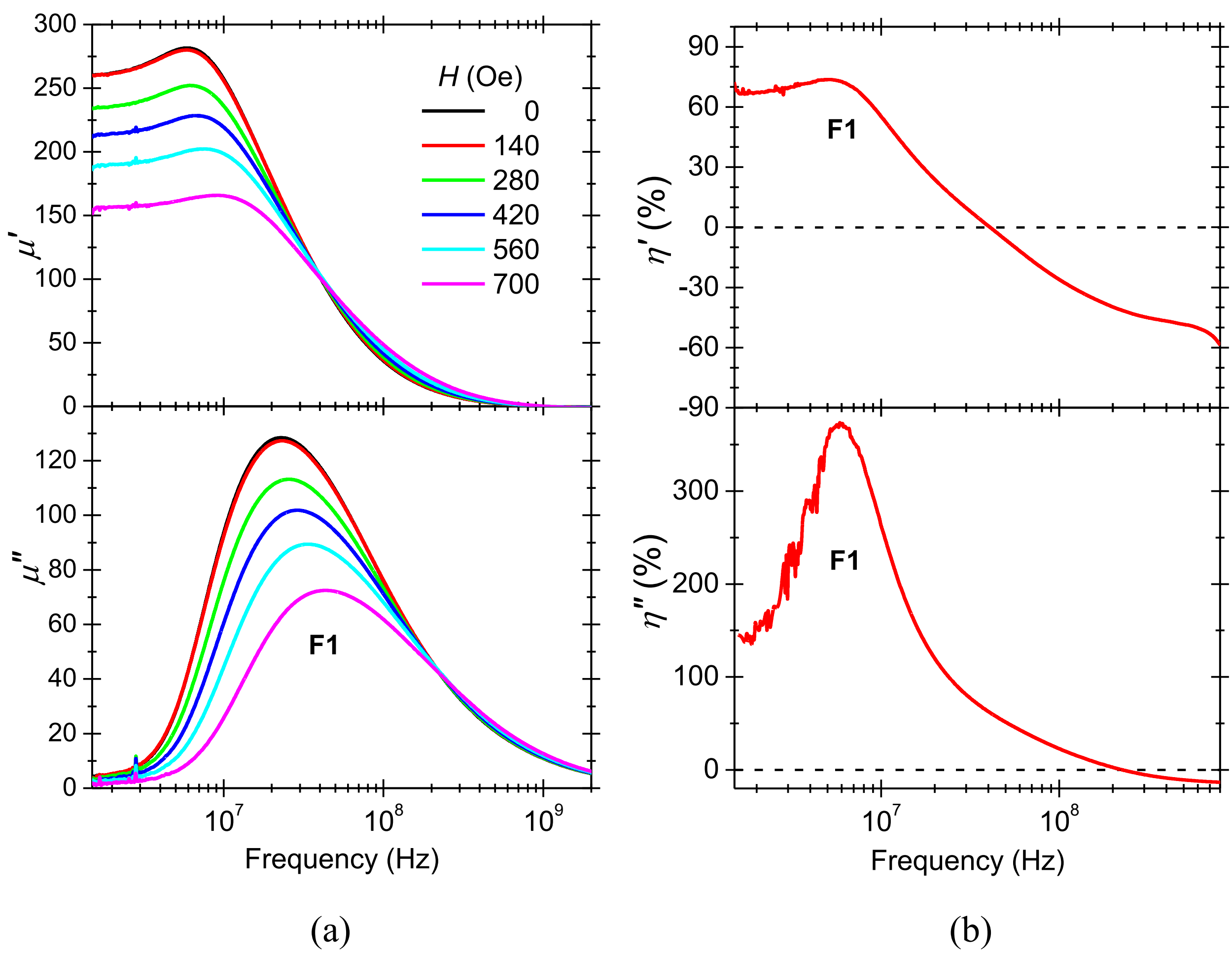

Fig. 6. Frequency dependences of the room-temperature (a) HF complex permeability at different magnetic bias fields and (b) HF magnetic nonlinearity of NZF ceramics. F1 is the magnetic excitation corresponding to the loss maximum.

The HF magnetic nonlinearity is high (Fig. 6b): $\eta'$ achieves values 60 % below 10 MHz and -40 % above 200 MHz, $\eta''$ achieves values above 120 % below 20 MHz (with a maximum 370 % at 6 MHz). The region of maximum nonlinearity is related to the F1 frequency at RT. Since we attribute F1 to the domain walls dynamics, the dynamics is strongly sensitive even to the weak bias magnetic field. So, NZF ceramics can be considered as a promising material for tunable HF devices.

## 5. **Microwave magnetic excitations**.

The weak bias magnetic field also influences the room temperature MW transmission $Tr(f)$, reflection $S_{11}(f)$ and reflection loss $RL(f)$ spectra of a microstrip line (MSL) loaded with the NZF sample (Fig. 7). Three frequency bands, where the $Tr(f)$ minima (transmission resonances), $S_{11}(f)$ anomalies and $RL(f)$ minima are well pronounced and field-dependent, can be attributed to three magnetic excitations (magnons) F2, F3, F4 corresponding to the collective spin dynamics of the domain bulk. Similar magnons were revealed in the U-type hexaferrite [28]. The high number of $RL(f)$ minima (Fig. 7b) is a result of the multiple EM wave reflections inside the sample, whose contribution is more seen in the $RL$ measurement [36].

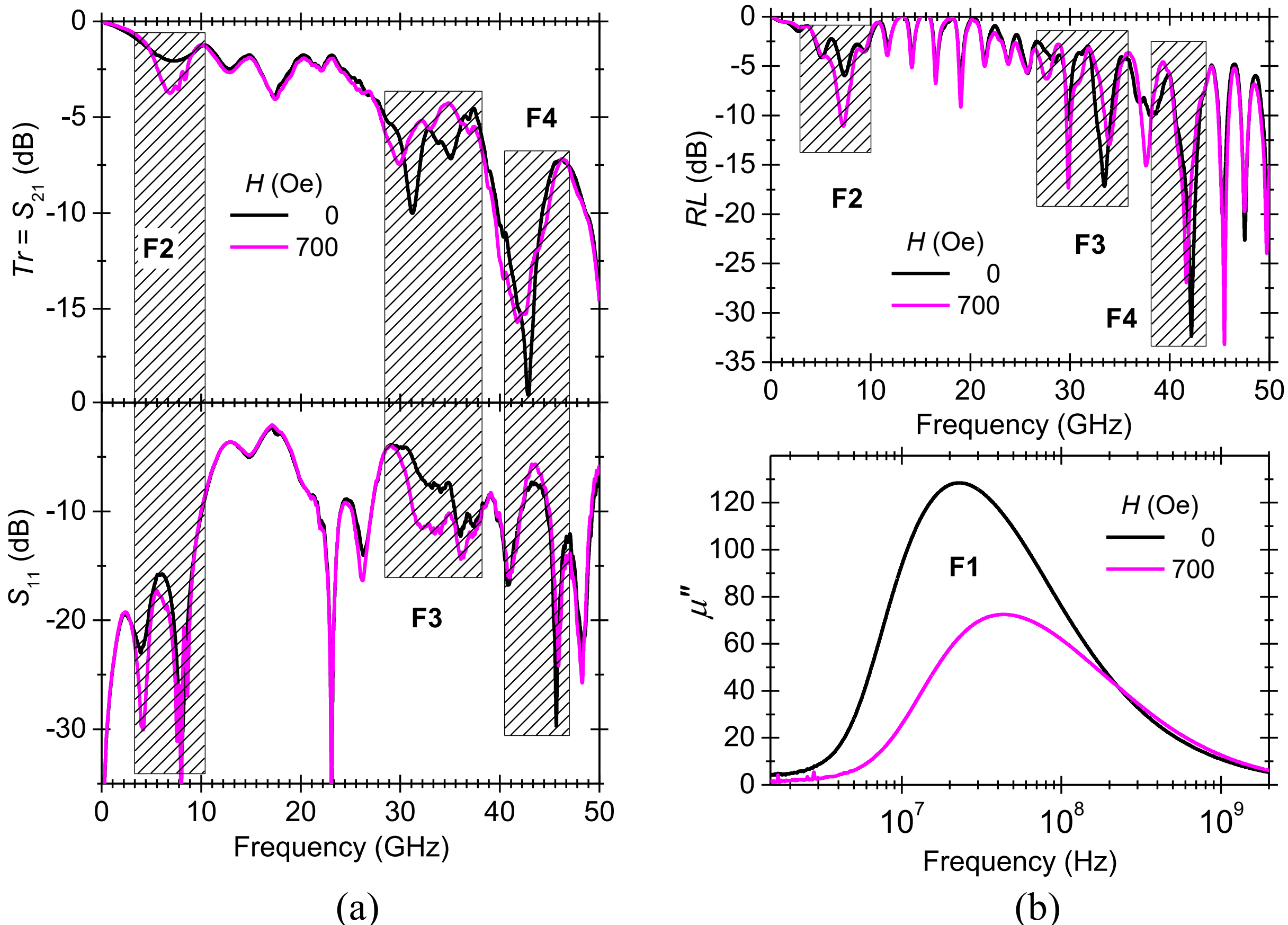

Fig. 7. Main magnetic excitations (F1 - F4) under the weak bias fields ($H = 0$ and $H = 700$ Oe) at RT: (a) MW spectra of the transmission and reflection coefficients, (b) MW reflection loss spectra and HF magnetic loss maximum. Dashed regions indicate the field-dependent frequency bands.

The multiple magnetic excitations are related to the complex magnetic structure of NZF consisting of the normal spinel and inverse spinel magnetic blocks [45] responsible for the formation of many magnon branches in the Brillouin zone [46]. Some of the magnons can be activated in the MW spectra. The lowest frequency magnon mode (F2) is the natural ferromagnetic resonance (FMR). Its frequency in a weak or zero magnetic field (~7 GHz) is usual for ferrites [28,47,48].

We suppose that the higher-frequency magnetic modes (F3 and F4) are other magnons of the exchange origin [46,49,50]. Let us note that the F3 and F4 modes are either split or include a few excitations with close frequencies (Fig. 7a). This can be explained by a difference in the magnetic moments' orientation in the different grains, i.e. by a difference in the magnetic anisotropy. On the other hand, we cannot exclude that F3 and F4 are magneto-dielectric resonance modes above FMR, even accounting for their magnetic field dependence and splitting [51,52]. Mixture of the magnon absorption and magneto-dielectric resonance is also possible. We will give a more precise definition to the F3 and F4 origin in our future studies.

To summarize, we revealed 4 magnetic excitations in NZF ceramics in a broad frequency range (10 MHz – 50 GHz) and attributed them to the domain wall and spin dynamics. Absorption bands corresponding to these excitations are shown together in Fig. 7b. Other anomalies in the transmission and reflection spectra, which are not field-dependent, can be attributed to the magneto-dielectric resonances (modes), like those observed in hexaferrites [28]. Both initial magnetic excitations and magneto-dielectric resonances contribute to the MW absorption and shielding efficiency of NZF.

## 6. Absorption and shielding efficiency.

The revealed magnetic excitations provide enhanced absorption of the electromagnetic waves. We estimate the MW absorption (*MA*) and shielding efficiency (*SE*) of NZF at room temperature in the frequency range 1 MHz – 2 GHz using spectra of the complex permeability (Fig. 4b) and frequency-independent complex dielectric permittivity $\varepsilon^* = \varepsilon' - i\varepsilon''$ with parameters $\varepsilon' = 13$ and $\varepsilon'' = 1.3$, which are in the frame of MW parameters usual for the ferrite ceramics [7,8,28]. Here we suppose that the possible dielectric dispersion, often observed at low frequencies and high temperatures [27,53] and related to the colossal permittivity effect caused by the inhomogeneous conductivity of ceramics [27,53–56], takes place in NZF mainly below the MHz frequency range. Then the complex permittivity can be considered as frequency independent between 1 MHz and 2 GHz. Of course, our assumption is rather rough, nevertheless it gives us the possibility to estimate the *MA* and *SE* caused by the magnetic properties of NZF. If the dielectric dispersion took place above 1 MHz, the related dielectric losses could only improve the estimated *MA* and *SE* parameters.

Based on the $\mu^*(f)$ and $\varepsilon^*(f)$ spectra below 2 GHz, we analyse interaction of the electromagnetic (EM) wave with the NZF slab (absorber) in a free space using two models presented in [37,38,57]: the EM wave transmission through the slab, accompanied with the partial reflection and absorption (Fig. 8a), and reflection from the slab backed by a perfect electric conductor (Fig. 8b). The EM wave is considered to be normal incident, the slab to be 2D-infinite.

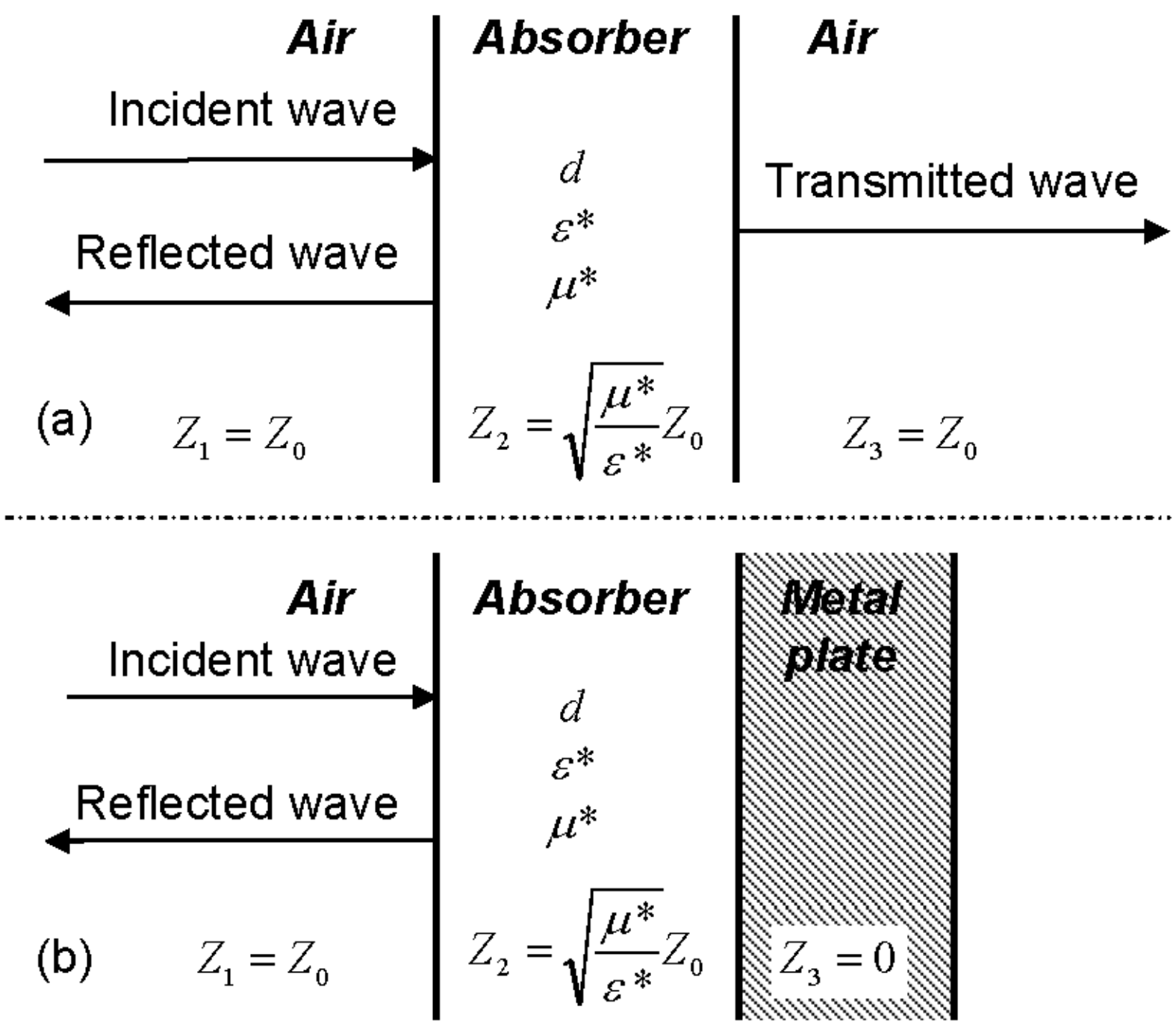

Fig. 8. Scheme of the electromagnetic wave interaction with the absorber in a free space: a) EM wave transmission through a slab between two semi-infinite media; b) EM wave reflection from the slab backed by a perfect electric conductor.

The characteristic impedance $Z_2$ and propagation constant $\gamma$ of the absorber are defined as

$$Z_2 = Z_0\sqrt{\frac{\mu^*}{\varepsilon^*}}, \qquad \gamma = \frac{i\omega\sqrt{\mu^*\varepsilon^*}}{c}, \tag{3}$$

where $Z_0 = 377\ \Omega$ is the characteristic impedance of free space, $c$ is the speed of light in free space, and $\omega$ is the angular frequency. The characteristic impedances $Z_1$ and $Z_3$ of the semi-infinite media are equal to $Z_0$ in the case of free space or air and $Z_3 = 0$ in the case of a metal plate.

Then in the model of EM wave transmission (Fig. 8a), the input wave impedance $Z_{\text{in}}$ at the slab surface can be defined as

$$Z_{in} = \sqrt{\frac{\mu^*}{\varepsilon^*}} Z_0 \frac{1 + \sqrt{\frac{\mu^*}{\varepsilon^*}}\tanh(\gamma d)}{\sqrt{\frac{\mu^*}{\varepsilon^*}} + \tanh(\gamma d)}, \tag{4}$$

where $d$ is the slab thickness. A modulus of the transmission coefficient $Tr$ can be calculated [37,38] as

$$Tr = 20\log\left|\frac{2Z_{in}}{Z_{in}+Z_0}\right|, \tag{5}$$

and the shielding efficiency is estimated as $SE = -Tr$.

In the model of EM wave reflection (Fig. 8b),

$$Z_{in} = \sqrt{\frac{\mu^*}{\varepsilon^*}} Z_0 \tanh(\gamma d). \tag{6}$$

Then the reflection loss ($RL$) and attenuation constant $\alpha$ (real part of the propagation constant $\gamma$) of the material, in neper/m, can be calculated [37,38]:

$$RL = 20\log\left|\frac{\sqrt{\frac{\mu^*}{\varepsilon^*}}\tanh(\gamma d) - 1}{\sqrt{\frac{\mu^*}{\varepsilon^*}}\tanh(\gamma d) + 1}\right|, \qquad \alpha = \mathrm{Re}(\gamma) = \mathrm{Re}\left(\frac{i\omega\sqrt{\mu^*\varepsilon^*}}{c}\right) \tag{7}$$

The reflection loss and attenuation constant characterize the MW absorption efficiency: $MA = -RL$. Spectra of the calculated absorption and shielding parameters are shown in Fig. 9 together with the F1 maximum of magnetic losses.

We cannot measure the $\mu^*(f)$ and $\varepsilon^*(f)$ spectra in the range from 2 GHz to 50 GHz, and then calculate the $SE$, $MA$ and $\alpha$ parameters of NZF in the free space. Instead, $SE$ and $MA$ were directly measured using the microstrip line loaded with the NZF sample as in the case of the U-hexaferrite ceramics [28]. The results are shown in Fig. 7 and in Figs 9a, 9b together with the calculated HF *SE* and *MA* parameters.

The HF free-space microwave absorption *MA* (Fig. 9b) is high at frequencies around the *RL*(*f*) minima which are defined by two factors: 1) magnetic and/or dielectric loss maximum, and 2) magnetodielectric (geometric) resonance related to the slab thickness, permittivity and permeability. In our case, the joint F1+R *RL*(*f*) minimum is defined by the F1 magnetic excitation at 10 – 100 MHz (no dielectric loss maxima are supposed) and geometric resonance R whose frequency can be tuned by the slab thickness. The same is valid for *SE* (Fig. 9a). The highest *MA* (lowest *RL* value) can be achieved when the F1 and R frequencies are close to each other. In our case, it corresponds to $d \sim 3 – 10$ mm. The frequency band of the highest *SE* (lowest *Tr* value in Fig. 7a) additionally depends on the slab reflection [57], therefore frequency bands of the highest *SE* and *MA* coincide (partially) only when the absorption prevails. It is just our case (compare Figs 9a and 9b): *SE* and *MA* bands are similar at $d \sim 3 – 10$ mm, and both are related to the domain wall excitation F1 and magnetodielectric (geometric) resonance R1.

Both *SE* and *MA* are dependent on *d*. Results for $d = 0.1 – 10$ mm allows to estimate the slab (absorber layer) optimum thickness for the selected frequency range. Nearly zero *SE* and *MA* values for absorber layers with $d = 0.1 – 0.3$ mm evidence that the layers are too thin in comparison with the HF free space wavelengths (> 300 mm below 1 GHz) and cannot effectively influence the electromagnetic wave propagation. Also, geometric resonances take place above 1 GHz. At frequences above 10 GHz, the same layers could be more effective. The layers with d = 3 – 10 mm are more optimal for the HF range.

The HF attenuation constant $\alpha$ of NZF (Fig. 9c) is higher than that of the dielectric-conductor poly(ethylene terephthalate) - carbon nanotube (PET-CNT) composites [58] and smaller than that of the giant-permittivity (In+Nb) co-doped CD-$TiO_2$ ceramics [56]. However, $\alpha$ of NZF rapidly increases with increasing frequency up to the CD-$TiO_2$ level at 1 GHz. This rapid increase of $\alpha$ is evidently caused by the $\mu''(f)$ maximum (shown in Fig. 9c) related to the magnetic domain wall dynamics.

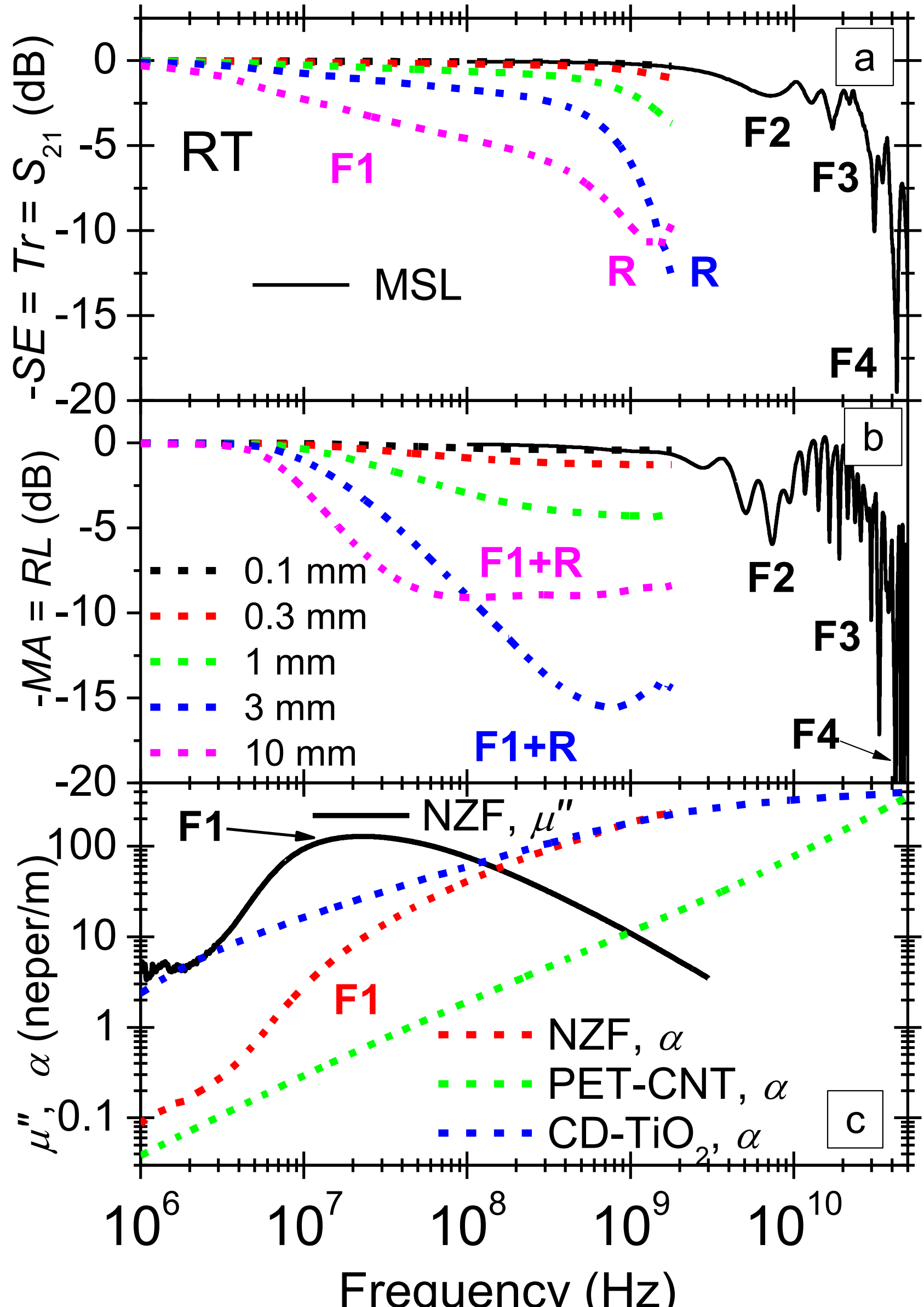

Fig. 9. Calculated spectra of the free space (a) shielding efficiency *SE*, (b) microwave absorption *MA* and (c) attenuation constant $\alpha$ of the NZF slabs with different thickness in the frequency range from 1 MHz to 2 GHz. *SE* and *MA* spectra of the NZF sample measured with the microstrip line up to 50 GHz are added to (a) and (b) respectively. Attenuation constant spectra of the PET-CNT [58] composites and (In+Nb) co-doped CD-$TiO_2$ ceramics [56] are added to (c) for comparison, as well as the NZF $\mu''(f)$ spectra. Main magnetic excitations (F1 - F4) and HF magnetodielectric (geometric) resonance R below FMR are related to the observed *Tr*(*f*) and *RL*(*f*) minima.

The MW *Tr*(*f*) and *RL*(*f*) minima above 2 GHz (Figs 9a and 9b) in the frequency bands of the F2, F3 and F4 magnons indicate also bands of the enhanced *SE* and *MA* in the transmission line (MSL in our case). The transmission line *SE* and *MA* values generally differ from the free-space ones, as well as the frequency bands of their maxima, due to the different impedance, sample and line geometry, type of the propagating waves, coupling between the wave and sample, etc. [36]. However, experimental data obtained in our MSL experiment allow us to estimate the free-space *SE* and *MA*, at least qualitatively. First, our MSL measuring cell ensures a single TEM mode propagation up to 50 GHz [35], i.e. the same wave type as propagates in the free space. Second, frequency bands of the MSL *Tr* and *RL* minima were shown to be mainly defined by the magnon absorption of the electromagnetic waves and therefore their frequencies mainly depend on the material parameters, not on the external conditions. Therefore, the F2, F3 and F4 frequency bands of the enhanced absorption and shielding presented in Fig. 9 are similar as in the free space.

In our MSL experiment, the coupling between sample and transmission line is rather low compared with models of the free space transmission through or reflection from the 2D-infinite slab (Fig. 8). Note (Figs 9a and 9b), that at frequencies where both the HF free space calculation and MSL experimental results are available (between 100 MHz and 2 GHz), the MSL experimental data correspond well to the lowest slab thickness (0.1 mm) in the free-space calculations. The lowest slab thickness corresponds to the lowest coupling. Therefore, one can suppose deeper *Tr*(*f*) and *RL*(*f*) minima in the F2, F3 and F4 bands in the case of higher coupling in the free space.

We consider that the natural FMR (the F2 magnon) can provide high absorption and shielding at frequencies below 10 GHz in the free space conditions of a real application. The electromagnetic wave absorption by the F3 and F4 magnetic modes can be high enough above 25 GHz, especially in the case of their mixing with the magneto-dielectric resonance modes. Even in the case of low coupling, the *SE* and *MA* values achieve 20 dB (Fig. 9). We also suppose that the attenuation constant $\alpha$ can increase above 2 GHz due to the FMR absorption contribution.

## 7. Conclusions

Single-phase $Ni_{0.5}Zn_{0.5}Fe_2O_4$ (NZF) nanoparticles with a cubic spinel structure and an average size of ~34 nm were synthesized and used to prepare uniform ceramics with ~10 μm grains. Our study of the magnetic properties of the NZF spinel ferrite ceramics over a broad frequency range from 1 MHz to 50 GHz revealed four magnetic excitations.

The high-frequency (HF) magnetic permeability of NZF ceramics, measured between 1 MHz and 2 GHz at temperatures from 100 K to 400 K, is high (~280 at 1 MHz and room temperature) and characterized by strong resonance-like dispersion at all temperatures with a broad, pronounced peak of the magnetic loss between 10 and 100 MHz. This dispersion is attributed to a HF magnetic excitation related to the magnetic domain wall dynamics. Based on the measured permeability spectra, HF absorption (*MA*), shielding efficiency (*SE*) and the attenuation constant were calculated using free-space models. Both *MA* and *SE* of NZF layers with an optimum thickness of ~3 mm were found to achieve 15 dB values below 1 GHz. The HF attenuation constant is high, reaching 100 neper/m, and is comparable to those of dielectric-conductor composites and giant permittivity materials. The HF absorption, shielding and attenuation of electromagnetic waves are shown to be related to the magnetic domain wall dynamics of NZF. The domain wall dynamics is strongly sensitive even to a weak bias magnetic field. It ensures high HF nonlinearity of the magnetic permeability and loss (at the level of ±50 %, or even more at some frequencies). Thus, the studied NZF ceramics can be considered a promising material for the development of HF absorbers and tunable HF devices.

In the microwave (MW) spectra of a microstrip line loaded with the NZF ceramic sample (between 2 and 50 GHz), we observed three frequency bands where transmission and reflection loss minima are well pronounced (down to -30 dB) and field-dependent. These frequency bands are attributed to three MW magnetic excitations (magnons) corresponding to the collective spin dynamics of the domain bulk. The lowest-frequency magnon (below 10 GHz) was attributed to the natural ferromagnetic resonance, while two other magnons are excited between 28 and 44 GHz. Other anomalies in the transmission and reflection loss spectra, which are not field-dependent, are attributed to magneto-dielectric resonances. Interaction of the revealed magnons and magneto-dielectric resonance modes with the electromagnetic wave could provide high absorption and shielding efficiency of NZF in the GHz range, including the 5G and 6G communication frequencies.

Our study contributes to a better understanding of the interaction between electromagnetic waves and the intrinsic magnetic dynamics of the $Ni_{0.5}Zn_{0.5}Fe_2O_4$ ceramics (as well as of other ferrites) related to domain walls, the natural ferromagnetic resonance and other microwave excitations.

## 8. Acknowledgments

This work has been supported by the Czech Science Foundation (Project No. 24-10791S) and by the project FERRMION (Project No. CZ.02.01.01/00/22_008/0004591), co-financed by the European Union and the Czech Ministry of Education, Youth and Sports.

The present work was also supported by the National Research Foundation of Ukraine in the framework of the project "Microwave devices based on resonant structures with metamaterial properties for the life protection and information security of Ukraine" (ID 2021.01/0030).